\documentclass[aps,prb,twocolumn,footinbib,showpacs]{revtex4}         
\usepackage{graphicx}

\newcommand{\ab}{{\em ab initio}\ }

\newcommand{\beg}{\begin{equation}}
\newcommand{\e}{\end{equation}}
\newcommand{\G}{\mathit{\Gamma}}
\newcommand{\half}{\frac{\scriptstyle 1}{\scriptstyle 2}}
\newcommand{\dhalf}{ {\textstyle\frac{1}{2}}  }
\newcommand{\Gseg}{G_{\mathrm{seg}} }

\newcommand{\Wsep}{W_{\mathrm{sep}} }
\newcommand{\As}{\textbf{A}$_\mathbf{s}$}
\newcommand{\Bs}{\textbf{B}$_\mathbf{s}$}
\newcommand{\Cs}{\textbf{C}$_\mathbf{s}$}
\newcommand{\Ds}{\textbf{D}$_\mathbf{s}$}
\newcommand{\Gcl}{$\mathcal{G}_\mathrm{cleav}$}
\newcommand{\Gdl}{$\mathcal{G}_\mathrm{disl}$}
\newcommand{\Gclm}{\mathcal{G}_\mathrm{cleav}}
\newcommand{\Gdlm}{\mathcal{G}_\mathrm{disl}}
\begin{document}

\title{Structural and chemical embrittlement of grain boundaries by impurities: a
general theory and first principles calculations for copper} 

\author{A. Y. Lozovoi, A. T. Paxton, and M. W. Finnis}
\altaffiliation[Present address: ]{Department of Materials, Imperial College of Science, Technology and Medicine, Exhibition Road, London SW7 2AZ, U.K.} 
\affiliation{Atomistic Simulation Centre, School of Mathematics and Physics,
Queen's University  Belfast, Belfast BT7 1NN, Northern Ireland, U.K.}  


\date{\today}

\begin{abstract}

First principles calculations of the $\Sigma 5$(310)[001]
symmetric tilt grain boundary in Cu with Bi, Na, and Ag
substitutional impurities provide evidence that in the
phenomenon of Bi embrittlement of Cu grain boundaries electronic
effects do not play a major role; on the contrary, the
embrittlement is mostly a structural or ``size'' effect. Na is
predicted to be nearly as good an embrittler as Bi, whereas Ag
does not embrittle the boundary in agreement with
experiment. While we reject the prevailing view that
``electronic'' effects ({\it i.e.,} charge transfer) are
responsible for embrittlement, we do not exclude the r\^ole of
chemistry. However numerical results show a striking equivalence
between the alkali metal Na and the semi metal Bi, small
differences being accounted for by their contrasting ``size'' and
``softness'' (defined here). In order to separate structural and
chemical effects unambiguously if not uniquely, we model the
embrittlement process by taking the system of grain boundary and
free surfaces through a sequence of precisely defined {\it
gedanken} processes; each of these representing a putative
mechanism. We thereby identify {\it three} mechanisms of
embrittlement by substitutional impurities, two of which survive
in the case of embrittlement or cohesion enhancement by
interstitials. Two of the three are purely structural and the
third contains both structural and chemical elements that by
their very nature cannot be further unravelled. We are able to
take the systems we study through each of these stages by
explicit computer simulations and assess the contribution of each
to the nett reduction in intergranular cohesion. The conclusion we
reach is that embrittlement by both Bi and Na is almost exclusively
structural in origin; that is, the embrittlement is a size
effect.

\end{abstract}

\maketitle

\section*{Introduction}

Impurity induced embrittlement of metals is an age old
phenomenon. For the case of bismuth in copper, it was apparently
first recorded in the scientific literature in the
1870s,\cite{Hampe} becoming a matter of both manufacturing
importance and scientific curiosity in the 20th century. As far
as we know the suggestion that segregation to grain boundaries was
the culprit was first made by Cyril S.~Smith in a comment to a
paper in 1929\cite{Blazey} and this explanation is now generally
accepted as the truth. Perhaps the most dramatic event resulting
from grain boundary segregation embrittlement was the failure of
the Hinkley Point `A' nuclear reactor in
1969.\cite{Kalderon,Gray} Given that fast, brittle failure is a
result of the segregation of insoluble impurities to grain
boundaries, the physical, microscopic origin of the loss of
cohesion is still a matter of intense debate and
controversy. And origins may indeed be different for different
metal and impurity combinations. At issue is the question, how
can impurities segregated to grain boundaries at concentrations
of a monolayer (ML) or less result in reduction in fracture toughness
amounting to a ductile to brittle transition?

Opinions are currently divided into two schools of thought
depending on whether it is believed that changes in the chemical
bonding {\em between host atoms} due to the presence of
impurities in their neighbourhood is the cause of the
embrittlement. The alternative to such {\it electronic}
mechanisms can loosely be described as a {\em size
effect}.\cite{Seah1,Seah2,Sutton}
The electronic mechanisms of grain boundary
embrittlement can in turn be divided into grain boundary
weakening and grain boundary stiffening arguments. In the former,
the impurities withdraw some electronic charge from neighbouring
host atoms which makes their cohesion across the boundary less
strong.\cite{MB}  In the latter, the impurity presence at the
boundary leads to the formation of strong directional bonds which
make it more difficult for dislocations to glide during
decohesion. This leads to local stress piling up in the vicinity
of the grain boundary, and the crystal falls apart in a brittle
manner.\cite{Haydock,Goodwin1,Goodwin2}

The other group of mechanisms, in which the effects of electronic
charge redistribution do not play a decisive role, has been
relatively little studied.\cite{braithwaite05} As we have recently
argued,\cite{rainer04} the well known effect of the embrittlement
of grain boundaries in copper by bismuth is a perfect
illustration of such 
embrittlement. Very briefly our
arguments were supported by three independent pieces of evidence
obtained during a study of the $\Sigma 19$a(331)[$\bar 1$10]
symmetric tilt grain boundary with segregated Bi.
Firstly, the electronic structure showed no evidence of
the ``white line'' in the electron energy loss near edge structure
(ELNES) reported experimentally to support the charge transfer
principle. (The existence of the white line has also been called
into question on experimental grounds.\cite{muller})  Secondly,
elastic constants of bulk Cu with impurities (Bi or Pb) strongly
depend on volume, but not on the presence of the
impurity. Thirdly, a Kanzaki analysis is consistent with a {\it
central force} model of Cu--Bi bonding, in which case the ease of 
dislocation emission from a crack tip should not be affected by 
the presence of Bi.
 
An opposite view was simultaneously expressed by Duscher~{\it et
al.},\cite{ruhle04} according to whom the role of Bi at a grain
boundary is to {\it donate} some of its electronic charge to the
neighbouring copper atoms rendering them ``zinc like'' and
resulting in the weakening of the Cu--Cu bonds across the
boundary. This point of view then falls into the class of
electronic mechanisms.
Their conclusion was based on a study of
the $\Sigma 5$(310)[001] symmetric tilt grain boundary with an
areal density of Bi atoms of 2.9~nm$^{-2}$ (approximately 0.25~ML
of Bi at the grain boundary plane). Evidence of charge transfer
between Bi and Cu taking place was based upon experimental
ELNES spectra. In addition, the authors of
ref.~[\onlinecite{ruhle04}] mention that the calculated
Bi induced expansion of the grain boundary is far too small to be
the reason for the grain boundary becoming weak if a ``size
effect'' argument were correct. On the other hand, our recent
arguments to support a size effect included the finding of a
large grain boundary expansion at the $\Sigma 19$a Bi segregated
grain boundary.\cite{rainer04}
   
We decided to address the controversy further by turning to the
the $\Sigma 5$ copper grain boundary so as to make a direct
comparison with the conclusions of Duscher~{\it et
al.}\cite{ruhle04} Our theoretical calculations presented for the
first time here, arrive at a very different conclusion.  The
grain boundary expansion turns out to be larger in our study and
we contend that the grain boundary studied in
ref.~[\onlinecite{ruhle04}] {\it was not embrittled}---indeed
these authors did not test this point either experimentally or
theoretically. We find that at least one monolayer of Bi at this
grain boundary is needed for embrittlement. Under those
circumstances the $\Sigma 5$ grain boundary is particularly
susceptible to Bi embrittlement.\cite{WangMRS}
Comparison of the effect of Bi with that of Na and Ag provide
further support for the principle that in the weakening of Cu grain
boundaries by Bi, charge redistribution plays a negligible role.

More importantly, in this paper we suggest a unified approach 
within which the effects of an impurity can 
be split into those related to mechanical distortion and to
chemical identities of impurity and host atoms. We then proceed
with our \ab data to demonstrate how these combine together
to explain the deterioration of grain boundary cohesion
for purely structural reasons.


The structure of this paper is as follows. We briefly summarise
the connection between fracture mechanics and thermodynamics of
decohesion in section~\ref{sec_theory}, in order to make clear
how the quantities we calculate relate to fracture
toughness. Section~\ref{sec_mech} offers a general approach 
as to how the various contributions which weaken or 
strengthen a given boundary can be quantified 
using explicit calculation of the changes in total energy during 
the stages of a thought experiment.
Technical details of the calculations, parameters used
and tests of the method are described in
section~\ref{sec_calc}. We present our results for the pure Cu
grain boundary and grain boundaries segregated with Bi, Na and Ag
in sections~\ref{sec_pure}, \ref{sec_bi} and~\ref{sec_nag}. 
Section~\ref{sec_how} aims to identify the dominant mechanism
responsible for grain boundary embrittlement by Bi and Na   
in terms of size, stiffness and chemical properties of the 
impurity atoms. The section follows the method outlined in 
section~\ref{sec_mech}. Our conclusions may be found in
section~\ref{sec_concl}.

\section{Thermodynamics of grain boundary embrittlement and
  relation to total energy in electronic structure calculations}
\label{sec_theory}

Grain boundary decohesion is a complex process involving
propagation of a crack under load and possibly simultaneous
emission of lattice or partial dislocations. Its proper
description requires atomistic simulations at the mesoscopic
level which is out of the reach of \textit{ab initio} techniques.
The approach that makes the problem manageable is due to Rice,
Thomson and Wang.\cite{rice1974,RiceWang} An atomically sharp
crack at a grain boundary will advance either in brittle or
ductile fashion depending on whether the minimal energy per unit
area of crack advance (energy release rate) required to emit a
single dislocation, \Gdl, is higher or lower than the
energy release rate associated with brittle cleavage of the
crystal along the grain boundary, \Gcl.\cite{rice1974,mason1979} 
If $\Gclm< \Gdlm$,
the crack remains atomically sharp during decohesion which is
therefore brittle, otherwise the system begins to emit
dislocations which blunt the crack.  In practice, \Gdl\ can
be accessed through the anisotropic theory of crystal defects;
for typical, pure Cu grain boundaries, \Gdl\ falls in the
region 1--2~J/m$^2$ (ref.~[\onlinecite{anderson86}]). Our task is
thereby reduced to determining whether segregation may reduce
\Gcl\ to a value significantly lower that this, hence
triggering a ductile to brittle transition.

As regards \Gcl, in the limit of perfectly elastic brittle
fracture, neglecting lattice trapping and energy dissipation
through phonon emission, it represents the reversible work needed
to convert unit area of grain boundary into twice that area of
free surface. The analysis is greatly complicated by impurity
segregation compared to fracture in pure materials. This is
because, whereas it may be taken that the impurity at the grain
boundary prior to fracture is in equilibrium with that in the
adjoining crystals, having an excess concentration $\G_{gb}$ per
unit area, the concentration profile at the surface cannot
generally be predicted {\it a priori}. As pointed out by
Seah,\cite{Seah1} and by Hirth and Rice\cite{hirth80} one may
imagine two limiting cases. In the first we suppose that the
growth of the crack is slow compared to the rate of concentration
equilibration between the surfaces produced and their underlying
bulk. In that case the grain boundary and surface excess
concentrations of impurity are at all times in accord with the
equilibrium adsorption isotherms and depend only on the
temperature and bulk concentrations, and we identify \Gcl\ with
the {\it work of adhesion},\cite{finnis1996} \beg
\label{Weq}
W_{\rm ad} = 2 \gamma_s - \gamma_{gb}. 
\e 
Here, $\gamma_{gb}$ and
$\gamma_{s}$ are the energies per unit area of the grain boundary
and surface whose impurity concentrations are such as to be in
equilibrium with the bulk. Equation~(\ref{Weq}) is known as the
Young--Dupr\'e equation\cite{SuttonBalluffi} when used to analyse
the equilibrium contact angle at a droplet on a surface. It is
relevant in fracture only if the impurity diffuses very rapidly
in the bulk, for example in instances of hydrogen embrittlement.
The second case is applicable here, namely that the growth of the
crack is so rapid that the impurity remaining at the surfaces has
no time to come into equilibrium with the underlying bulk. Indeed
at best some surface diffusion may occur as the crack is opening
and one may suppose a {\it local equilibrium}\cite{hirth80} is
achieved in the surface layer of host and impurity
atoms.\footnote{In the context of our {\it ab initio}
calculations this is achieved by atomistic relaxation and by
exploring a limited number of surface adsorption sites---assuming
that these lie in the top layer of the surface.} In that case one
identifies \Gcl\ with the reversible {\it work of
separation}\cite{finnis1996} under the constraint that bulk
diffusion is suppressed. As now the number of particles, and not
their chemical potentials, are fixed during the separation we
obtain 
\beg
\label{Wsep}
\Gclm = \Wsep(\G_{gb}) = \frac{1}{A} \left\{2 G_s(\dhalf\G_{gb}) -
G_{gb}(\G_{gb}) \right\},
\e
where $A$ is the surface area, $G$ is the excess Gibbs free energy of 
a representative piece of material containing either the grain boundary
(before decohesion) or the surface (after decohesion), $\G_{gb}$ 
is the impurity excess at the grain boundary before cleavage, and we 
assume that the impurity is evenly distributed between 
the two newly created surfaces. The excess in Eq.~(\ref{Wsep}) is 
taken with respect to the adjoining bulk phases. 

To express equation~(\ref{Wsep}) in terms of surface and grain
boundary energies one can follow the analysis of Hirth and
Rice.\cite{hirth80} Under the condition that the bulk impurity
concentration is dilute 
we may write
\beg
\label{Wneq}
\Wsep(\G_{gb}) = 2 \gamma_s(\dhalf\G_{gb}) - \gamma_{gb} +
(\mu_i^\prime - \mu_i) \G_{gb}.  
\e 
This may be useful in
revealing two differences between the ``slow'' and ``fast''
limiting cases. In finding equation~(\ref{Wneq}) it is necessary
to regard the surface atomic layer as a thermodynamic system,
uncoupled from the underlying bulk. The impurity atoms have a
chemical potential $\mu_i^\prime$, different from their chemical
potential $\mu_i$ in the bulk and at the grain boundary prior to
fracture.  One can then interpret $\gamma_s(\half\G_{gb})$ as the
surface energy of a crystal whose bulk composition is such that
its equilibrium surface excess is $\half\G_{gb}$. Secondly, the
final term in equation~(\ref{Wneq}) may be interpreted as the
energy released to the impurity atoms as a result of the change
in chemical potential on finding themselves out of equilibrium
with the underlying bulk. This illustrates the point that one
would always expect the work of adhesion to be smaller than the
work of separation.

Both of the limiting cases embodied in equations~(\ref{Weq})
and~(\ref{Wsep}), may alternatively be formulated in terms of the
work of adhesion or separation of the equivalent grain boundary
in pure material \textbf{A}, say, $\Wsep(\mathbf{A})$, plus terms
accounting for the presence of impurity.\cite{Seah1,Seah2}. 
In the case of ``fast'' fracture, if one neglects a small
contribution to $\Wsep$ associated with the change of
configurational entropy when a grain boundary with impurity
splits into surfaces, equation~~(\ref{Wsep}) becomes:\cite{Seah2}
\begin{equation}
\label{Wseg}
\Gclm = \Wsep(\mathbf{B}) = \Wsep(\mathbf{A}) - \Gseg(\mathbf{B_s})+\Gseg(\mathbf{B})\,,
\end{equation}
where \textbf{B} denotes material with segregant, as in figure~(\ref{fig_paths}), and
$\Gseg(\mathbf{B_s})$ and $\Gseg(\mathbf{B})$ are the segregation free energies 
per unit area of surface and grain boundary respectively. 
Equation~(\ref{Wseg}) may be
interpreted to imply that an impurity which segregates more
readily to a free surface than to a grain boundary will cause a
reduction in work of separation, or conversely will lead to grain
boundary cohesion enhancement.\cite{RiceWang} 
It is worth repeating that a
reduction in work of separation does not necessarily imply {\it
embrittlement\/} which in the simplified picture advocated here
requires that $\Gclm < \Gdlm$.

In our \ab supercell calculations the effect of temperature is neglected
hence we approximate the Gibbs free energies $G_{gb}$ and $G_{s}$ in (\ref{Wsep})  
with total energies $E_{gb}$ and $E_{s}$ of the supercells 
containing the grain boundary and the surface, respectively. Correspondingly, 
the segregation free energies in (\ref{Wseg}) are estimated as the energy required to 
remove all the impurity from the interface and distribute it in the bulk. 


The assumption that impurity atoms distribute evenly between the
surfaces in equation~(\ref{Wsep}) may be removed
in a straightforward way.  Modification to any arbitrary
distribution is allowed as long as the surface excesses
sum up to that at the grain boundary.  For example, if one of the
surfaces stays bare, $2 G_s(\half\G_{gb})$ in equation~(\ref{Wsep})
must be replaced with $G_s(\G_{gb}) + G_s(0)$.  In principle, one
should choose the distribution that minimises the sum of the
``surface'' Gibbs free energies $G_s$ and therefore the work of
separation (\ref{Wsep}). In our study, we consider two limiting
cases having either equal distribution, or all impurities
residing on one of the two surfaces. The difference is generally
rather modest; experiment usually shows an even distribution of
Bi between the surfaces.\cite{PowellMykura73}

\section{Structural and chemical embrittlement}
\label{sec_mech}

In this section we define  concepts of ``structural
embrittlement'' and ``chemical embrittlement'' 
precisely. While there is no unique definition that can
unambiguously separate structural from chemical effects, we seek
a definition that is practical and could be used to rationalise
results for many different systems.  For this reason a general
definition should be unambiguous and independent of the
components, including both interstitial and substitutional
impurities. For future applications we formulate such a general
definition, although the calculations described in this paper are
for substitutional impurities only. Furthermore, some
modification to the following discussion will need to be made to
describe the ``slow'' fracture of equation~(\ref{Weq}).

\subsection{Intermediate configurations}
\label{sec_points}

\begin{figure}
\begin{center}
\includegraphics[scale=0.7,angle=0, trim=0 0 0 0]{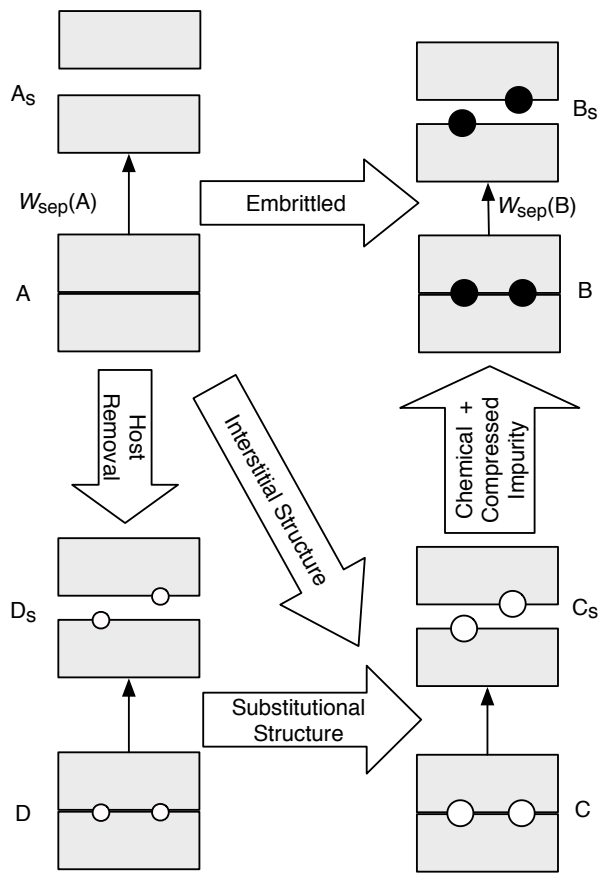}
\end{center}
\caption{General scheme for discussion of embrittlement
mechanisms due to boundary weakening by impurity segregation. 
{\bf A} and {\bf B} denote the fully relaxed equilibrium 
unsegregated and segregated grain boundaries, respectively. 
Boundary {\bf C} is created by substituting impurity atoms
in {\bf B} (black spheres) with vacancies (white spheres) 
without further relaxation, and {\bf D} is created from {\bf A} 
in the same way. \textbf{A$_s$--D$_s$} denote the  
free surfaces into which  grain boundaries \textbf{A--D} cleave. 
Path \textbf{A$\to$D$\to$C$\to$B} refers to substitutional impurities,
whereas path \textbf{A$\to$C$\to$B} applies to interstitial impurities. 
}
\label{fig_paths}
\end{figure}

In order to proceed we need to
define a number of hypothetical reference systems (see figure~\ref{fig_paths}), 
which we denote \textbf{A}, \textbf{B}, \textbf{C} and \textbf{D}. Each system
is in the form of a grain boundary of unit area and separated surfaces, the
latter being distinguished from the grain boundary by the suffix~$\mathbf{s}$. 

These systems refer to a given host and impurity combination, a given grain
boundary and a given distribution of impurity atoms at the boundary.  A
system comprising separated surfaces should have exactly the same total
numbers of atoms of each species as the corresponding grain boundary,
so that bulk quantities will cancel out of all the relevant energy
differences, as  described by works of separation. To avoid a complicated treatment of
ensembles that would be required to deal with more concentrated solutions, we assume that on
the length scale of a supercell\footnote{
Although the discussion in this section is of a general nature, we employ the language
of computer simulations. Thus ``supercell'' refers to a representative piece of the system 
periodically repeated in space, ``relaxation'' denotes minimisation of the internal energy 
of the system with respect to atomic positions, and so on. (For details see section~\ref{sec_calc}).}   
the bulk can be treated as pure host. The systems are defined as follows:\\
\\
\textbf{A}: The relaxed grain boundary of the pure
 host. \\ 
\As: The relaxed, separated surfaces corresponding to~\textbf{A}. \\
\textbf{B}: The relaxed grain boundary of the host with segregated impurity atoms. \\
\Bs: The relaxed, separated surfaces corresponding to \textbf{B}. \\
\textbf{C}: The same atomic structure as \textbf{B}, with impurity atoms removed from the
impurity sites defined in \textbf{B} but no further relaxation. We refer to the imprint of these absent
impurities as ``ghosts'' since they define forces that maintain the host atoms in
a distorted structure, but make no other direct contribution to the energy. For substitutional
impurities a ghost is a more or less strained vacancy. For interstitial impurities a
ghost is a centre of strain in the host lattice without missing atoms.\\ 
\Cs: The
same atomic structure as \Bs, with impurity atoms removed from the impurity sites defined in
\Bs\ but no further relaxation.\\  
\textbf{D}: This system is only relevant for substitutional
impurities such as those discussed in the rest of this paper. It is the same structure as \textbf{A},
with host atoms removed from the impurity sites defined in \textbf{B} but no further relaxation. \\
\Ds: The same atomic structure as \As, with
host atoms removed from the impurity sites defined in \Bs\ but no further relaxation.\\

Some points of clarification may help to visualise these systems.
Note first that \textbf{A}, \As, \textbf{B} and \Bs\ label fully relaxed structures, whereas \textbf{C},
\Cs, \textbf{D} and \Ds\ label notional structures we set up in the computer using the atomic
positions found for the relaxed systems. The works of
separation for each system are defined as the differences in total energy, for example for system \textbf{A}:
\begin{equation}
\Wsep(\mathbf{A}) = E(\mathbf{A}_s) - E(\mathbf{A})\,.
\end{equation}
Although we would prefer $E$ to mean the Gibbs free energy, for
simplicity of calculation we use as a surrogate the total energy at $0\,$K (see section~\ref{sec_theory}). 
A more exact calculation could be made in principle, to include phonon free
energies in the harmonic approximation. 

If the impurity is detrimental to cohesion, the work of separation of our embrittled boundary, \textbf{B}, is less
than the work of separation of the clean boundary, 
$\Wsep(\mathbf{A})-\Wsep(\mathbf{B}) > 0$. As summarised in section~\ref{sec_theory}, this picture of
grain boundary weakening is equivalent to the statement that surface segregation
is energetically preferred to grain boundary segregation, that is, from equation~(\ref{Wseg}), 
\begin{equation}
\Wsep(\mathbf{A})-\Wsep(\mathbf{B}) = \Gseg(\mathbf{B}_s) - \Gseg(\mathbf{B}) > 0 \,.
\end{equation}
These formulations of the difference show how the bulk energies, including the energy
contribution of dissolved impurities, cancel out. We are now in a position to discuss what we
mean by a ``size effect'' or ``structural effect'' as opposed to a ``chemical effect''.

\subsection{Size effects and chemical effects}
\label{sec_steps}

Leaving aside for the moment chemical effects, we might regard a misfitting impurity in the
bulk as storing energy mechanically, like a compressed or stretched spring.  The stored energy
is fully released when the impurity segregates to a surface, but only partially released when
the impurity segregates to a grain boundary. This would be a simple mechanical picture of the
origin of the embrittlement. However, it is unsatisfactory without specifying what happens to
the stored energy. Consider the case of an oversized and therefore substitutional impurity. If
it found a sufficiently loose site at the grain boundary, it could eliminate the strain energy
and lower the grain boundary energy by segregating there. In this case there would be no further
reduction in stored energy on cleavage. The net effect would therefore be an \emph{increase} in
$\Wsep$.  While the stored energy in an oversized, compressed impurity is likely to be a
driving force for getting it to a boundary, what matters for embrittlement is how much of this
energy is \emph{stored} in the boundary, to be released on cleavage. It is stored in three
ways:
({\it i\/})
as broken host--host bonds, because the impurity substitutes for the host;
({\it ii\/})
as stretched host-host bonds; and
({\it iii\/})
in the compressed or expanded impurity atoms themselves.

If the impurity is interstitial, the first of these is irrelevant, since we suppose that no
host--host bonds in the boundary are broken by the presence of the impurity. The first two
contributions are illustrated in figure~\ref{fig_paths} by the paths labelled ``Host
Removal (HR) $\rightarrow$ Substitutional Structure (SS)'' and ``Interstitial Structure (IS)'' respectively.
System \textbf{D} was introduced solely for the purpose of quantifying the magnitude of the
substitutional effect, by removing the effects of bond-stretching. Accordingly we define the
following contributions to the path from \textbf{A} to \textbf{C} for substitutional structures:
\beg\label{HR}
\mathrm{HR} = \Wsep(\mathbf{D}) - \Wsep(\mathbf{A})
\e
\begin{equation}\label{SS}
\mathrm{SS} = \Wsep(\mathbf{C}) - \Wsep(\mathbf{D})\, ,
\end{equation}
and for interstitial impurities
$$
\mathrm{IS} = \Wsep(\mathbf{C}) - \Wsep(\mathbf{A})\,.
$$
The final step around the diagram in figure~\ref{fig_paths},
labelled ``Chemical + Compressed Impurity (CC)'' takes the system
from \textbf{C} to \textbf{B}: \beg\label{CC} \mathrm{CC} =
\Wsep(\mathbf{B}) - \Wsep(\mathbf{C})\, .  \e This is where the
chemistry of the impurity comes into play. Notice that in this
step the host atoms are in the same positions with and without
the impurity, so no further energy change is associated with the
energy stored in the host lattice.  However, if we are thinking
mechanically, there may still be some energy released by volume
relaxation of the impurities in going from the grain boundaries
to the surfaces. We might expect this to be more significant when
the grain boundary concentration is low, in which case there are
still some host-host bonds holding the impurities in a state of
compression, compared to the case of say a monolayer of
segregation, in which the local volumes of the impurities are
much more relaxed. However, it does not seem possible or useful
in the step from \textbf{C} to \textbf{B} to try to separate this
from the chemical contribution, which contains the energy of the
impurity-host bonds that are broken.  Note that our IS and CC
mechanisms are respectively the ``mechanical'' and ``chemical''
contributions identified in refs.~[\onlinecite{geng99,janisch03}]
in the context of interstitial impurities.

What would be the likely effect of the contributions just
considered on the work of separation?  The HR mechanism should
\textit{decrease} $\Wsep$ since removal of a host atom cuts more
bonds at a grain boundary than at a surface. Conversely, the CC
step is expected to \textit{increase} $\Wsep$, although the
necessity to squeeze an oversized impurity while inserting it
into a grain boundary acts so as to reduce the effect. The
reduction should be most pronounced for tight grain boundary
sites or large and soft impurity species. (By ``large'' and
``soft'' we imply a measure such as found in table~\ref{tbl_fcc},
below.) Finally, the SS step must \textit{decrease} $\Wsep$ for
an oversized impurity, as the only way for such an impurity to
accommodate itself at the grain boundary is to push the grains
apart, whereas at a free surface it may protrude into the vacuum
without significantly distorting its surrounding host atoms.

The interstitial impurity case reduces to the competition between
only two mechanisms (IS and CC); the result can be of either
sign.\cite{geng99,janisch03,braithwaite05} Interestingly, the IS
mechanism tends to \textit{strengthen} the boundary for small
impurities (H, B, C, N, and O) as they distort free surfaces more
than grain boundaries by occupying subsurface sites, whereas the
effect of the CC mechanism can be either positive or negative
depending on the nature of chemical bonds formed between
impurities and host atoms.\cite{geng99,janisch03} Indeed, in
ref~[\onlinecite{janisch03}] it is found that B and C enhance the
cohesion of grain boundaries in Mo, while N and O reduce the work
of separation; the contribution of the CC mechanism also acts to
lower $\Wsep$ if the impurity is C, O or N, while only for B does
the CC mechanism lead to cohesion enhancement at their grain
boundary in Mo.

We show in what follows that one can carry the system around the
\textit{gedanken} path in figure~\ref{fig_paths} explicitly by
calculating the quantities defined by
equations~(\ref{HR})--(\ref{CC}) and hence identify the
magnitudes of the ``size'' and ``chemical'' contributions to
grain boundary decohesion.

\section{Calculation details}
\label{sec_calc}

\begin{table*}
\caption{Double basis set parameters used in the present study: atomic radius, $r_a$, smoothing radii $r_{sm}$, and
energetic parameters $-\kappa^2$ for each angular momentum $\ell$. Radii $r_a$ are chosen so that 
atomic spheres would not overlap by more than few percent in all configurations considered.}
\begin{tabular}{lccccccccc}
\hline
\hline
\multicolumn{1}{c}{Species} & \multicolumn{1}{c}{$r_a$,}
                            & \multicolumn{4}{c}{$r_{sm}/r_a$}
                            & \multicolumn{4}{c}{$-\kappa^2$, Ry}\\
\cline{3-10}
                &  Bohr    & $\ell=0$ & $\ell=1$ & $\ell=2$ &$\ell=3$ 
                            & $\ell=0$ & $\ell=1$ & $\ell=2$ &$\ell=3$ \\
\hline
Cu & 2.238 & 0.761 & 0.667 & 0.442 &       & --0.600 & --0.400 & --0.400 &         \\
   &       & 0.761 & 0.831 & 0.417 &       & --1.000 & --1.000 & --1.000 &         \\
Bi & 2.450 & 0.801 & 0.756 & 0.486 & 0.800 & --0.911 & --0.374 & --0.400 & --0.400 \\
   &       & 0.801 & 0.756 & 0.486 & 0.800 & --3.000 &         & --3.000 & --1.000 \\
Na & 2.450 & 0.569 & 0.582 & 0.469 &       & --1.636 & --0.400 & --0.400 &         \\
   &       & 0.569 & 0.582 & 0.469 &       & --2.000 & --2.000 & --2.000 &         \\
Ag & 2.450 & 0.528 & 0.551 & 0.408 &       & --1.031 & --0.400 & --0.555 &         \\
   &       & 0.528 & 0.551 & 0.408 &       & --3.000 & --3.000 & --3.000 &         \\
\hline
\hline
\end{tabular}
\label{tbl_basis}
\end{table*}

We have calculated the atomic and electronic structure of the
$\Sigma 5$ grain boundary in copper from first principles 
using the full potential LMTO method as implemented in
the NFP code.\cite{nfp} In this code, wavefunctions and charge
density are expanded in smooth Hankel functions having an energy
$-\kappa^2$ and a smoothing radius $r_{sm}$ defined such that
beyond $r_{sm}$ smooth and singular Hankel functions are not
appreciably different.  For each atom type we employ two such
envelope functions for each angular momentum quantum number up to
some $\ell_{max}$. These are augmented inside atomic spheres with
numerical solutions of the radial Schr\"{o}dinger equation;
all relativistic effects except spin orbit coupling are included,
the core is permitted to relax to self consistency but is treated
as spherical. In the {\it interstitial region\/} outside the atomic
spheres, the Poisson equation is solved analytically in the
smooth Hankel basis and the exchange--correlation energy and
potential are obtained through a real space grid representation. 
We use the local density approximation (LDA) in the
parameterisation of von Barth and Hedin, modified by Moruzzi {\it
et al.}\cite{vBH,MJW} Basis function parameters and atomic sphere
radii used in our study, are shown in table~\ref{tbl_basis}. For
Bi we additionally included a localised orbital to allow the Bi
$5d$ semi core to be treated as a valence state in addition to
the $6d$ valence state.

\begin{table*}
\caption{Bulk properties of Cu and Bi: lattice constant, $a_0$, rhombohedral angle, $\alpha$, and internal coordinate, $u$,
of the equilibrium crystal structure (fcc for Cu, and A7 for Bi), structural energy differences, $\Delta E_\beta$ with
respect to that structure, bulk modulus, $B$, and elastic constants, $C^\prime=(c_{11}-c_{12})/2$ and $c_{44}$, 
obtained in the present study, other scalar relativistic (unless indicated otherwise) LDA calculations, and experiment.}
\begin{tabular}{lcccccc}
\hline
\hline
\multicolumn{1}{c}{} 
& \multicolumn{3}{c}{Cu}
& \multicolumn{3}{c}{Bi}\\
\cline{2-7}
& \multicolumn{1}{c}{our LDA} & \multicolumn{1}{c}{other LDA}& \multicolumn{1}{c}{Exp.} 
& \multicolumn{1}{c}{our LDA} & \multicolumn{1}{c}{other LDA}& \multicolumn{1}{c}{Exp.} \\
\hline
$a_0$, \AA             & 3.5270 & 3.53$^a$, 3.52$^b$,   & 3.6146\cite{Smithells} 
                      & 4.6211 & 4.62$^a$, 4.528$^e$   & 4.7460\cite{degtyareva04}  \\    
                      &        & 3.58$^c$, 3.52$^d$    &                            \\ 
$\alpha$              &        &                       &          
                      & 58.25$^\circ$ & 58.23$^\circ$$^a$, 58.93$^\circ$$^e$ & 57.23$^\circ$\cite{degtyareva04}\\    
$u$                   &        &                       &          
                      & 0.2370 & 0.238$^a$, 0.238$^e$  & 0.2341\cite{degtyareva04}  \\    
$\Delta E_{fcc}$, mRy &  0     &  0                    &  0                      
                      & 11.4   &                       &                             \\    
$\Delta E_{bcc}$, mRy & 2.7    & 2.9$^a$, 3.2$^f$      &                         
                      & 7.0    &                       &                             \\    
$\Delta E_{sc}$, mRy  & 39.4   & 40.0$^f$              &                         
                      & 1.5    &                       &                             \\    
$B$, GPa              & 185.9  & 190$^a$, 192$^b$,   & 137\cite{Kittel}        
                      & 43.4   & 46$^a$,  51$^a$       & 38.2\cite{degtyareva04}      \\    
                      &        & 153$^c$, 190$^d$  &                 \\      
$C^\prime$, GPa       & 31.2   & 35$^a$, 27.2$^c$, 25$^d$       &  23.5\cite{Simmons}                     
                      &        &                       &                             \\    
$c_{44}$, GPa         & 98.9   & 99$^a$, 86$^c$, 80$^d$&           75\cite{Simmons}
                      &        &                       &                             \\    
\hline
\hline
\end{tabular}
\newline
$^a$ Ultra soft pseudopotentials, Refs.~[\onlinecite{rainer04,rainer}]. \\
$^b$ LAPW, ref.~[\onlinecite{khein1995b}].                              \\   
$^c$ Full potential LMTO, non-relativistic, ref.~[\onlinecite{kraft1993}].\\    
$^d$ Full potential LAPW, ref.~[\onlinecite{mehl1996}].   \\   
$^e$ Norm conserving pseudopotentials, ref.~[\onlinecite{gonze1988}].   \\   
$^f$ LAPW, ref.~[\onlinecite{mishin2001}].   \\   
\label{tbl_bicu}
\end{table*}

\begin{figure}[ht]
\begin{center}
\includegraphics[scale=0.85,angle=0]{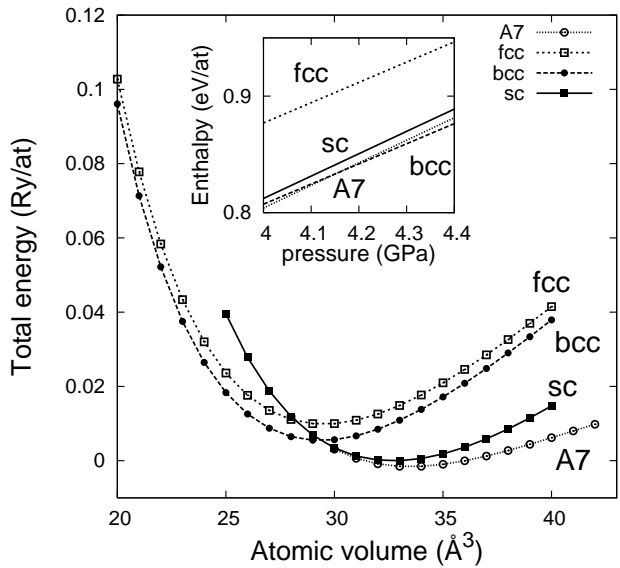}
\end{center}
\caption{Energy--volume curves for bulk Bi. A fragment of the 
respective enthalpy--pressure plot near the bcc--A7 crossover is shown
in the inset. The energy and enthalpy are measured relative to those of
Bi(A7) at zero pressure.}
\label{fig_bi}
\end{figure}

Some bulk properties of Cu and Bi obtained in our LDA
calculations are listed in table~\ref{tbl_bicu} where they are
compared with experiment and other LDA calculations.
Figure~\ref{fig_bi} shows calculated energy--volume curves for bulk
Bi.  In agreement with experiment, the rhombohedral A7 structure
has the lowest energy. With pressure increasing, the A7 curve
gradually merges into the simple cubic curve, the rhombohedral
angle steadily increasing towards 60$^\circ$ and the internal
coordinate $u$ tending to its symmetrical value of 0.25. However,
at a pressure of 4.2~GPa the A7 curve is intersected by the bcc
curve. In experiment, the bcc structure becomes stable above
7.7~GPa, although there are two more phases, Bi--II and Bi--III,
which appear between A7 and bcc.\cite{degtyareva04} These are
rather complex and were not considered in our study.

The grain boundary was represented in our calculations with a
periodic supercell containing two identical grain boundaries
separated by nineteen (310) layers (38 atoms). The equilibrium
width of the grain boundary was determined by stretching the
supercell along the interface normal and finding the minimum of
the total energy of the resulting relaxed supercell as a function
of the stretch. The lateral dimensions of the supercell were
fixed by the theoretical equilibrium lattice parameter of the
bulk fcc copper (table~\ref{tbl_bicu}).  Atomic positions were
fully relaxed in each of the supercells considered until the
forces dropped below 1 mRy/Bohr. 
No finite temperature
effects were taken into account. As the equilibrium structure of
the $\Sigma 5$(310)[001] grain boundary in Cu is well known from
previous theoretical \cite{sorensen2000} and experimental
\cite{ruhle04} studies, in our calculations we have not attempted
to further optimise the atomic structure of the grain boundary by
shifting the adjacent grains laterally. To minimise systematic
errors in calculating the work of separation, the energy of free
surfaces was obtained using the same supercell, in which about
half of the copper layers were replaced with the vacuum so that
no grain boundary remains. For the same reason, we filled the
whole supercell with atoms (so called ``space filling slab''
geometry) for the reference bulk calculations. For calculating
grain boundary and surface excess free energies,
equation~(\ref{Wsep}), at different coverages and hence different
stoichiometries of the slabs we employ standard
formulas\cite{finnis98,lozovoi01} taking the chemical potentials of the
components from bulk calculations.

Calculations were done for impurity atoms (Bi, Na, or Ag)
replacing 0, 1 (``loose'' site, see~figure~\ref{fig_gb}), and 2
atoms at the grain boundary plane, which we refer to as 0, 0.5
ML, and 1 ML coverage, respectively. In addition, we considered
the grain boundary with Bi at 0.25 ML coverage by doubling the
supercell along the [001] tilt axis and replacing every
\textit{other} Cu atom in the ``loose'' position with Bi.  In all
supercell calculations, a 16$\times$10$\times$2 Monkhorst--Pack
mesh of $k$-points was employed with a Methfessel--Paxton
broadening factor \cite{methfessel1989} of 5~mRy. To represent
the smooth interstitial electronic density,\cite{nfp} a
24$\times$38$\times$152 real space mesh was employed.  In
practical terms, a significant speed up of the calculations was
achieved by performing the relaxation first with a somewhat
reduced set of parameters (8$\times$4$\times$2 $k$-point mesh and
a 18$\times$25$\times$96 real space mesh) employing the criterion
that forces were stabilised within \mbox{1--2} mRy/Bohr, and then
further optimising the positions using the finer set of meshes. A typical
difference between the work of separation and grain boundary
expansion obtained with either set of parameters, 0.05 J/m$^2$
and 2 pm, respectively, can be taken as a ``pessimistic''
estimation of the convergence of our calculations with respect to
the number of $k$-points and to the size of the real space mesh.

Fully relaxed calculations of the impurity atoms in Cu bulk were
done using 32 atom cubic cells with an 8$\times$8$\times$8
$k$-point mesh and a 48$\times$48$\times$48 real space mesh.

\section{Pure copper $\Sigma 5$ grain boundary}
\label{sec_pure}

\begin{figure*}[ht]
\begin{center}
\includegraphics[scale=0.65,angle=0]{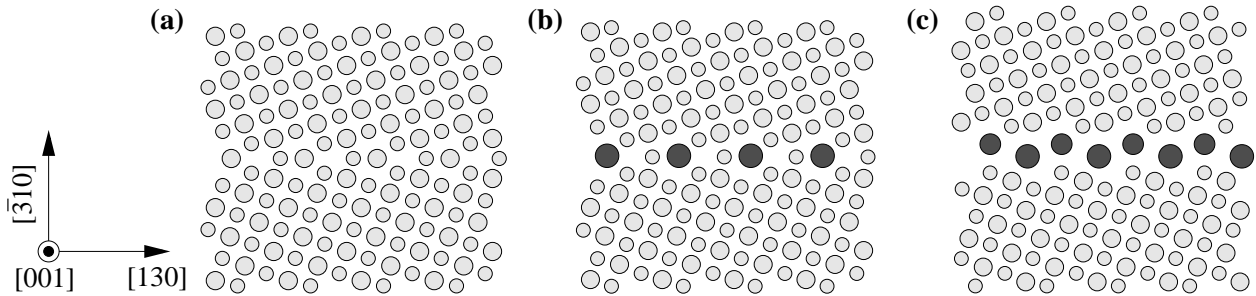}
\end{center}
\caption{Relaxed atomic structure of the Cu $\Sigma 5$(310)[001]
grain boundary: (a) Pure copper; (b) with 0.5 ML of Bi; and (c)
with 1 ML of Bi at the grain boundary plane. Notice that in the
pure grain boundary (a) there are two inequivalent atomic
positions in the boundary plane. We refer to these as ``loose''
(the larger circle) and ``tight'' (the smaller circle) to
illustrate the average distance of the nearest
neighbours. Naturally we expect a larger atom preferentially to
select the loose site to occupy.}
\label{fig_gb}
\end{figure*}

\begin{table}
\caption{The energy of the (310) surface, $\gamma_s$, and of the
$\Sigma 5$(310)[001] grain boundary, $\gamma_{gb}$, the grain
boundary width, $w$ (grain boundary excess volume per unit area),
and the work of separation, $\Wsep$, in pure Cu.  Experimental
data and the results of other calculations are given in
parentheses.  For comparison, the results for the $\Sigma
19$a(331)[$\bar 1$10] grain boundary and (331)
surface\cite{rainer04} are also shown.}
\begin{tabular}{lcc}
\hline
\hline
\multicolumn{1}{c}{Pure Cu} & \multicolumn{1}{c}{$\Sigma 5$}      
                            & \multicolumn{1}{c}{$\Sigma 19$a$^d$}\\
\hline

$w$, \AA               &  0.28                        &  0.22\\  
$\gamma_s$, J/m$^2$    &  2.21   (1.77$^a$)           &  2.07 \\
$\gamma_{gb}$, J/m$^2$ &  1.07  (0.980$^b$,0.888$^c$) &  1.04 \\
$\Wsep$, J/m$^2$     &  3.35                        &  3.10 \\
\hline
\hline
\end{tabular}
\newline
$^a$ Exp. polycryst., ref.~[\onlinecite{tyson77}]. \\
$^b$ Finnis-Sinclair many-body potentials, ref.~[\onlinecite{alber1999}].\\ 
$^c$ EAM, ref.~[\onlinecite{sorensen2000}]. 
$^d$ Ultra soft pseudopotentials, ref.~[\onlinecite{rainer04}]. \\
\label{tbl_cu}
\end{table}
      
The atomic structure of the relaxed $\Sigma 5$(310)[001]
symmetric tilt grain boundary in pure Cu obtained in the present
study, is shown in figure~\ref{fig_gb}a, as viewed along the [001]
direction. It contains two alternating (001) planes separated by
half the lattice constant, $a_0$, shown with larger and smaller
circles. It can be imagined as a mirror related twin boundary
structure in which one of the two (310) boundary planes in
either grain bordering the mirror plane is removed, while the
other one collapses into the grain boundary plane.  The same
structure can be obtained by arranging a suitable lateral shift
for the upper half of the original twin with subsequent
relaxation.\cite{sorensen2000} One ends up with two 
non--equivalent atoms at the grain boundary plane, one of which being
in the ``loose'' environment and the other being in the ``tight''
environment.  The former provides a natural substitutional site
for an oversized impurity. A recent experimental STEM image of
the grain boundary\cite{ruhle04} is very similar to that depicted
in figure~\ref{fig_gb}a.  The calculated grain boundary energy
(table~\ref{tbl_cu}) is in good agreement with semi empirical
data.  In addition, the $\Sigma 5$ results listed in the table
are close to those of \ab calculations
obtained for the $\Sigma 19$a(331)[$\bar 1$10] grain boundary in pure
Cu.\cite{rainer04} The works of separation of both grain
boundaries are sufficiently large to ensure that the boundaries
are ductile according to the criterion $\Wsep > \Gdlm
\simeq$ 1--2~J/m$^2$ as they should be.

\begin{figure}[ht]
\begin{center}
\includegraphics[scale=0.8,angle=0]{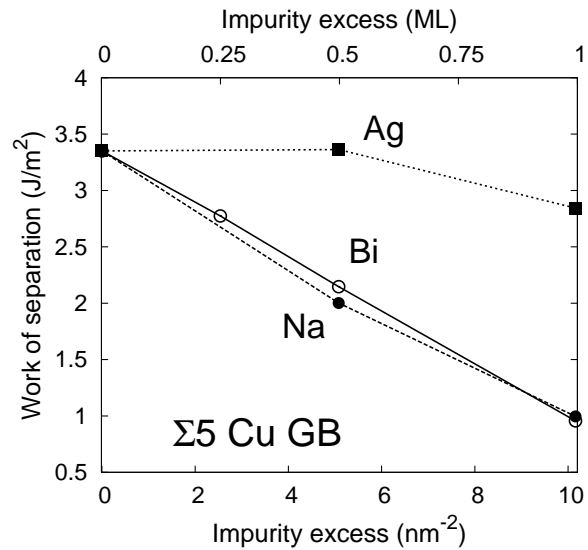}
\end{center}
\caption{Dependence of the work of separation on the impurity excess at the $\Sigma 5$(310)[001] Cu grain boundary:
Bi (open circles), Na (filled circles), and Ag (filled squares).}
\label{fig_wsep}
\end{figure}

\begin{figure}[ht]
\begin{center}
\includegraphics[scale=0.8,angle=0]{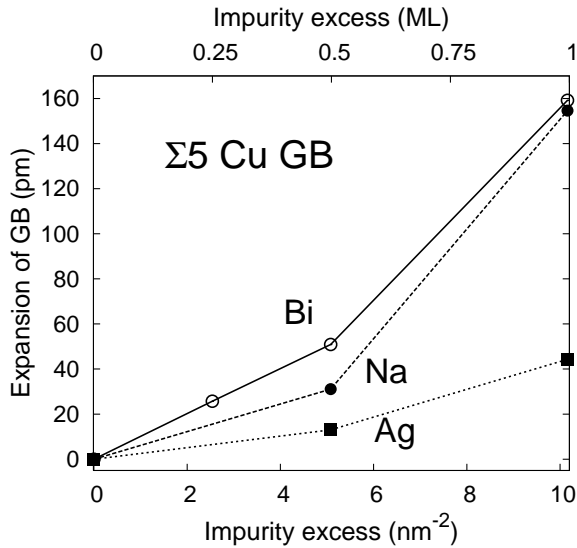}
\end{center}
\caption{Same as figure~\protect\ref{fig_wsep} but for the grain boundary expansion.}
\label{fig_expan}
\end{figure}

\section{Copper $\Sigma 5$ grain boundary with bismuth}
\label{sec_bi}

\begin{table*}
\caption{Grain boundary expansion and work of separation of
the $\Sigma 5$(310)[001] Cu grain boundary with impurities.  Grain
boundary coverage of 0.5~ML and surface coverage of 1~ML
correspond to an areal density of 5.08 impurity atoms per nm$^2$.
}
\begin{tabular}{lcccccccc}
\hline
\hline
                           & \multicolumn{1}{c}{Cu}      
                           & \multicolumn{3}{c}{Bi} & \multicolumn{2}{c}{Ag} 
                           & \multicolumn{2}{c}{Na} \\
\cline{2-9}
Impurity excess, ML        &   --  & 0.25 & 0.5  &     1 & 0.5 & 1 & 0.5 & 1 \\
\hline
GB expansion $\delta$, pm  &   0  &  25.7 &  50.9 & 159.2 &  13.0 & 44.5   &  31.1  & 154.6  \\
Segregation energy, eV:                                                                      \\
to the $(310)$ surface   &   0  &       & 3.11& 3.01& 0.77& 0.77 & 3.00 & 2.70 \\
to the grain boundary      &   0  & 1.69& 1.63& 1.54& 0.79& 0.46 & 1.34 & 1.26 \\ 
$\Wsep$, J/m$^2$:                                                                          \\
into 0.5~+~0.5 surfaces     & 3.35 &      &  2.15 &  0.96 &  3.37 & 2.84   &   2.00 & 1.00   \\
into 1~+~0  surfaces        & 3.35 & 2.77 &  2.23 &  1.85 &  3.36 &        &   2.24 &        \\
\hline
\hline
\end{tabular}
\label{tbl_pt}
\end{table*}

The relaxed atomic positions of the
$\Sigma 5$ grain boundary with 0.5 ML of Bi are shown in
figure~\ref{fig_gb}b; those for the 0.25 ML of Bi are very similar
except that the grain boundary expansion is about half as large
(see table~\ref{tbl_pt}).

We define the grain boundary expansion as the difference per
grain boundary between the lengths of supercell with and without
impurity atoms, noting that supercells contain the same number of
atoms since our impurities are {\it substitutional}.  In view of
the large discrepancy in the grain boundary expansion of 50.9~pm
for the 0.5~ML Bi~segregated grain boundary obtained in the
present study compared with the calculation of Duscher~{\it
et~al.},\cite{ruhle04} namely 7--10~pm, using the ultra soft
pseudopotentials (USPP) within the VASP code,\cite{vasp1-3} we
have additionally computed the expansion of one of the $\Sigma
19$a boundaries with 1 ML of Bi, for which we already knew the
USPP result obtained with the VASP code.\cite{rainer} As the NFP
and VASP expansions were found to agree within 6\% (102~pm and
108~pm, respectively) we can safely conclude that the grain
boundary expansions listed in table~\ref{tbl_pt} are the correct
results within the local density approximation. 

The work of separation and the grain boundary expansion in table~\ref{tbl_pt} 
are also plotted in figs.~\ref{fig_wsep} and~\ref{fig_expan} against  
the Bi excess at the boundary. 
Up to 0.5~ML of Bi, both quantities comfortably fall on a straight
line, meaning that while Bi atoms fill the ``loose'' sites  
their effect is additive. 


Although the work of separation decreases as Bi progressively
substitutes Cu at the ``loose'' sites of the grain boundary, even
at 0.5~ML it remains larger than \Gdl, which implies that
up to that coverage, {\it the boundary is ductile}. This will
have been the situation in the work of Duscher~{\it et
al.}\cite{ruhle04} From experiment it is known that the
$\Sigma 5$ grain boundary is \textit{severely} embrittled by
Bi.\cite{WangMRS,li1995b} We therefore proceeded to add more Bi
at the grain boundary by replacing Cu both at the ``loose'' and
``tight'' sites (1~ML). In this calculation we did not fully
optimise the grain boundary structure with respect to rigid
lateral shifts of the grains, however we removed the mirror
symmetry constraint so that the grains could move during the
atomic relaxation if they wished. As can be seen in
figure~\ref{fig_gb}c, the relaxed structure of the grain boundary
does show a significant lateral displacement of the grains; in
addition the Bi layer substantially buckles giving rise to large
increase of the grain boundary width. The buckling of the grain
boundary plane is due to large Bi atoms being too close to each
other, perhaps the structure shown in figure~\ref{fig_gb}c can be
considered as a precursor of the grain boundary faceting which is
often detected at Cu grain boundaries with
Bi.\cite{donald79,sigle2002,straumal06} Specifically in the
$\Sigma 5$ grain boundary, some ``roughening'' has been
reported.\cite{alber1999}

The work of separation of the Cu grain boundary with 1 ML of Bi
drops below 1~J/m$^2$ which we take as an indication that
the threshold concentration of the impurity required to embrittle
the grain boundary, has been reached. Note in passing that the
way in which impurity atoms distribute themselves between the
surfaces after cleavage turns out to be of little importance;
the case of equal distribution (denoted as ``0.5~+~0.5'' in
table~\ref{tbl_pt}), is marginally lower in energy than the case
when all Bi atoms reside on one of the surfaces (``1~+~0'' in
table~\ref{tbl_pt}). The exception is the ``1~+~0'' separation of
the 1 ML grain boundary demonstrating a sizeable increase of
$\Wsep$.

The segregation energies of Bi to the grain boundary depend
weakly on Bi concentration there (which is just another way to
say that the $\Wsep({\mit\Gamma})$ dependence is linear) and are
significantly smaller than the segregation energies to the free
(310) surface, as is expected for an embrittling impurity (see
Sec.~\ref{sec_theory}).
Our grain boundary segregation energies are comparable to those obtained
for the $\Sigma 19$a boundary,\cite{rainer04} namely 1.32~eV
for 0.5 ML and 1.58~eV for 1 ML per impurity atom. The experimentally reported
Gibbs segregation free energy to the $\Sigma 5$ boundary at $T =
700^{\circ}$C is $G_{\rm seg} =$
0.92$\pm$0.11~eV---markedly smaller than our theoretical
values. This was, however, the largest segregation
energy among the all grain boundaries studied in
ref.~[\onlinecite{alber1999}].  The difference may well be due to
the effect of vibrational entropy. McLean\cite{McLean} argued
that the segregation to grain boundaries of solute atoms
sufficiently larger or smaller than the host atoms, results in
the atomic packing becoming denser at the grain boundary, hence
the vibrational entropy of the grain boundary region may
decrease.  Decrease of the vibrational entropy has been recently
predicted for Ag segregating to the $\Sigma 5$ grain boundary
plane in Cu.\cite{berthier02}

\section{Copper $\Sigma 5$ grain boundary with sodium and silver}
\label{sec_nag}

If Bi embrittlement is a purely structural effect which has
nothing to do with Bi's chemical identity, then the same
behaviour is expected from other elements which, say, are of
about the same size as Bi and are not engaged in any strong
chemical bonding with the host atoms (see section~\ref{sec_mech}
for a more detailed discussion of structural embrittlement).
In ref.~[\onlinecite{rainer04}] it was shown that Pb behaves very
similarly to Bi. One can argue, however, that since Bi and
Pb are neighbours in the Periodic Table, such similarity is
indeed expected. We therefore decided to pick an element as
different to Bi as possible, from which only structural
embrittlement is expected. For this reason, we chose the alkali
metal Na. Like Bi it does not form any stable compounds with Cu
and has vanishingly small solubility.\cite{massalski}
Indeed, the calculated enthalpy of solution of Na in fcc Cu is as
large as that of Bi (table~\ref{tbl_bulk}).  In addition, the
lattice constants of Na and Bi, both calculated in a hypothetical
fcc structure, are close to each other and much larger than the
lattice constant of Cu (table~\ref{tbl_fcc}).  Na, however, is a
much softer species than Bi as is exemplified by its smaller bulk
modulus (table~\ref{tbl_fcc}) and solute dilation volume
(table~\ref{tbl_bulk}).

\begin{table}
\caption{Enthalpy of solution, $H_s$, and the relative dilation volume, $\Omega_d/\Omega_0$, 
per impurity atom 
of Bi, Ag, and Na in bulk fcc Cu. $\Omega_d$ is defined as the difference of the equilibrium
volumes of the cell with and without impurity at constant number of atoms, and $\Omega_0$ is 
the equilibrium atomic volume of pure Cu. The last column gives the formation enthalpy (in eV) and 
formation volume (in units of $\Omega_0$) of a vacancy in bulk Cu.} 
\begin{tabular}{llcccccc}
\hline
\hline
\multicolumn{2}{l}{Impurity:}       
                           & \multicolumn{1}{c}{Bi} 
                           & \multicolumn{1}{c}{Ag} 
                           & \multicolumn{1}{c}{Na}   
                           & \multicolumn{1}{c}{Vac} \\
\hline
$H_s$, eV:   
&    this work             &    1.66    & 0.50              & 1.73  & 1.27 \\
&    other calc.           &    1.7$^a$ &0.83$^b$           &       & 1.33$^{b,f}$, 1.29$^g$ \\
&    exp.                  &            &0.39$^c$           &       & 1.28$^h$\\
$\Omega_d/\Omega_0$:  
&    this work             &    1.40    & 0.54              & 0.67  & 0.64 \\
&    other calc.           &    1.3$^a$ &0.64$^b$, 0.52$^d$ &       & 0.70$^{b,d}$\\
&    exp.                  &            &0.45$^e$, 0.58$^e$ &       & 0.75$^h$\\
\hline
\hline
\end{tabular}
\newline
$^a$ Ultra soft pseudopotentials, Refs.~[\onlinecite{rainer04,rainer}]. \\
$^b$ LSGF, unrelaxed, ref.~[\onlinecite{korzhavyi99r}]. \\
$^c$ Exp., ref.~[\onlinecite{Hultgren}]. \\
$^d$ FP-KKR, ref.~[\onlinecite{papanikolaou97}]. \\
$^e$ Exp., ref.~[\onlinecite{pearson}]. 
$^f$ FP-LMTO, ref.~[\onlinecite{korhonen95}]. \\
$^g$ FP-KKR, ref.~[\onlinecite{drittler91}]. \\
$^h$ Exp., ref.~[\onlinecite{Ehrhart91}]. \\
\label{tbl_bulk}
\end{table}

\begin{table}
\caption{Lattice constant, $a_0$, and bulk modulus, $B$, of the impurity species calculated in 
bulk fcc structure. These data may be used as a measure of the relative
``size'' and ``softness'' of atoms.}
\begin{tabular}{llccccc}
\hline
\hline
\multicolumn{1}{l}{Impurity:} & \multicolumn{1}{c}{Cu}      
                           & \multicolumn{1}{c}{Bi} & \multicolumn{1}{c}{Ag} 
                           & \multicolumn{1}{c}{Na} \\   
\hline
$a_0$, \AA  & 3.5270 & 4.9048& 4.0218 & 4.9779 \\
$B$, GPa    &  185.9 & 64.7  & 135.3  & 15.8   \\
\hline
\hline
\end{tabular}
\label{tbl_fcc}
\end{table}

The other element that we have chosen for comparison is Ag for it
is known that silver \textit{does not} embrittle Cu. The lattice
constant of fcc Ag is nevertheless larger than that of Cu, and
the heat of solution of Ag in Cu is positive, although these are
not as extreme as for either Bi or Na. The experimental Ag--Cu phase
diagram shows noticeable solubility of Ag in Cu (up to 5\%), but
without the formation of stable compounds.\cite{massalski}

The grain boundary calculations with Na and Ag were done in
the same way as for Bi.  The expansion of the grain boundary with
0.5 ML of Na is smaller compared to Bi, however the grain
boundary with 1.0 ML of Na expands by almost the same amount as
with Bi (see table~\ref{tbl_pt} and figure~\ref{fig_expan}). This
can be understood assuming that Na is a softer species than Bi,
but of slightly larger size (see table~\ref{tbl_fcc}) and taking
into account that the fall off of the mutual attraction between
Cu grains with distance is faster than linear. In all other
respects Bi and Na behave strikingly similarly (see
table~\ref{tbl_pt} and figure~\ref{fig_wsep}). In particular, we
observe that Na renders the grain boundary brittle at 1 ML
coverage. This is our prediction, as no experimental
studies of Na at Cu grain boundaries seem to exist. However,
there are indications that Na can embrittle Cu in the context of
liquid metal embrittlement.\cite{LME}

\begin{table}
\caption{Works of separation $\Wsep$ (in J/m$^2$) corresponding to configurations \textbf{A}--\textbf{D} in figure~\ref{fig_paths}
for impurity coverages of 0.5 and 1 ML. Configuration \textbf{A}
(pure boundary) of course does not depend on impurity coverage.} 
\begin{tabular}{lcccc|ccc}
\hline
\hline
\multicolumn{1}{c}{Impurity} &
& \multicolumn{3}{c|}{0.5 ML}      
& \multicolumn{3}{c}{1 ML} \\
\cline{3-8}
& \multicolumn{1}{c}{\textbf{A}}
& \multicolumn{1}{c}{\textbf{B}}
& \multicolumn{1}{c}{\textbf{C}}
& \multicolumn{1}{c|}{\textbf{D}}
& \multicolumn{1}{c}{\textbf{B}}
& \multicolumn{1}{c}{\textbf{C}}
& \multicolumn{1}{c}{\textbf{D}}\\
\hline
Bi & 3.35 & 2.15 & 1.81 & 2.20  &  0.96 & 0.13  & 1.38 \\
Na & 3.35 & 2.00 & 1.97 & 2.20  &  1.00 & 0.14  & 1.38 \\
Ag & 3.35 & 3.37 & 2.17 & 2.20  &  2.84 & 1.01  & 1.38 \\
\hline
\hline
\end{tabular}
\label{tbl_points}
\end{table}

\begin{table*}
\caption{Impurity segregation energies (in eV per impurity atom)
related to $\Wsep$ in table~\ref{tbl_points} by virtue of
equation~(\ref{Wseg}). The sign convention is such that a
positive energy means that the impurity wants to segregate. Note that
the impurity in configurations \textbf{C} and \textbf{D}
is actually a vacancy, both in the bulk and at the
interface (see section~\ref{sec_mech}).}
\begin{tabular}{lcccccc|cccccc}
\hline
\hline
\multicolumn{1}{c}{Impurity}
&\multicolumn{6}{c|}{0.5 ML} 
&\multicolumn{6}{c}{1 ML} \\
\cline{2-13}
& \multicolumn{2}{c}{\textbf{B}} & \multicolumn{2}{c}{\textbf{C}} & \multicolumn{2}{c|}{\textbf{D}} 
& \multicolumn{2}{c}{\textbf{B}} & \multicolumn{2}{c}{\textbf{C}} & \multicolumn{2}{c}{\textbf{D}}\\
\cline{2-13}
& \multicolumn{1}{c}{surf} & \multicolumn{1}{c}{gb}
& \multicolumn{1}{c}{surf} & \multicolumn{1}{c}{gb}
& \multicolumn{1}{c}{surf} & \multicolumn{1}{c|}{gb} 
& \multicolumn{1}{c}{surf} & \multicolumn{1}{c}{gb}
& \multicolumn{1}{c}{surf} & \multicolumn{1}{c}{gb}
& \multicolumn{1}{c}{surf} & \multicolumn{1}{c}{gb} \\
\hline
%
Bi & 3.11 & 1.63 &   1.18& $-$0.71 &  1.20& $-$0.22 &  3.01 & 1.54 &   1.23& $-$0.75 &  1.22 & 0.02 \\
Na & 3.00 & 1.34 &   1.24& $-$0.46 &  1.20& $-$0.22 &  2.70 & 1.26 &   1.27& $-$0.70 &  1.22 & 0.02 \\
Ag & 0.77 & 0.79 &   1.21& $-$0.25 &  1.20& $-$0.22 &  0.77 & 0.46 &   1.25& $-$0.19 &  1.22 & 0.02 \\
\hline
\hline
\end{tabular}
\label{tbl_eseg}
\end{table*}

The similarity between Bi and Na provides 
evidence that charge transfer effects are of negligible
importance. If for example the major cause of the embrittlement
were weakening of Cu--Cu bonds due to transfer of electrons {\it
from} Bi {\it to} Cu as proposed in ref.~[\onlinecite{ruhle04}],
then a free electron metal such as Na would have led to even
stronger embrittlement, say, at 1~ML coverage where the grain
boundary expansion for Bi and Na is the same. This is not
observed in our calculations.

Silver, on the other hand, does not bring about any significant
decrease of $\Wsep$ even at 1~ML (table~\ref{tbl_pt} and
figure~\ref{fig_wsep}). In fact, at 0.5~ML coverage it even
slightly strengthens the boundary, despite a small expansion of
the latter. Such a peculiar situation arises because Ag atoms
turn out to be of just the right size to replace Cu at the
``loose'' sites.  The resulting denser packing at the grain
boundary thus compensates the replacement of Cu--Cu bonds with
slightly weaker Cu--Ag bonds. Interestingly the energy of Ag
segregation to the grain boundary decreases in going from 0.5~ML
to 1~ML: this simply reflects the attractive nature of the 
``loose'' site to the slightly larger Ag atom.

\section{How does bismuth embrittle copper? }
\label{sec_how}


\begin{table}
\caption{The changes in $\Wsep$ (in J/m$^2$) corresponding to
Substitutional Structure (SS), Host Removal (HR), and Chemical +
Compressed Impurity (CC) mechanisms as defined in
figure~\ref{fig_paths}.  Negative numbers indicate a decrease in
$\Wsep$.  For Bi and Na, the HR mechanism provides the largest
contribution.  However, at 1 ML its relative importance
diminishes with the SS mechanism contributing nearly half of the
overall decrease of $\Wsep$. The CC mechanism always acts so as to
increase $\Wsep$.  For Ag, the dominant contributions come from
the HR and CC mechanisms which are nearly compensating. (This is
an almost trivial result. If we employed Cu as the ``impurity''
then HR and CC would both be large and would exactly cancel,
while SS would be zero.)
}
\begin{tabular}{lcccc|cccc}
\hline
\hline
\multicolumn{1}{c}{Impurity} 
                           & \multicolumn{4}{c|}{0.5 ML}      
                           & \multicolumn{4}{c}{1 ML} \\
\cline{2-9}
& \multicolumn{1}{c}{Total}
& \multicolumn{1}{c}{SS}
& \multicolumn{1}{c}{HR}
& \multicolumn{1}{c|}{CC}
& \multicolumn{1}{c}{Total}
& \multicolumn{1}{c}{SS}
& \multicolumn{1}{c}{HR}
& \multicolumn{1}{c}{CC} \\
\hline
%
%
Bi &  $-$1.20 &  $-$0.39 &  $-$1.15 & 0.34    &    $-$2.39  & $-$1.25  & $-$1.97 & 0.83     \\ 
Na &  $-$1.35 &  $-$0.23 &  $-$1.15 & 0.03    &    $-$2.35  & $-$1.25  & $-$1.97 & 0.86     \\
Ag &     0.02 &  $-$0.04 &  $-$1.15 & 1.20    &    $-$0.51  & $-$0.37  & $-$1.97 & 1.83   \\
\hline
\hline
\end{tabular}
\label{tbl_steps}
\end{table}


In accordance with the analysis presented in Sec.~\ref{sec_mech},
we have created intermediate configurations \textbf{C} and \textbf{D}
defined in Sec.~\ref{sec_points} which together with fully
relaxed pure (\textbf{A}) and segregated systems (\textbf{B})
form a complete set of data required.  The works of separation
for each of these configurations corresponding to 0.5 ML and 1 ML
coverages are listed in table~\ref{tbl_points}. These works of
separation are further split into surface and grain boundary
contributions in terms of the segregation energies in
table~\ref{tbl_eseg}.  Finally, table~\ref{tbl_steps} shows the
changes of $\Wsep$ invoked by transitions between the
configurations \textbf{A}--\textbf{D} that give the contributions
of the Substitutional Structure (SS), Host Removal (HR), and
Chemical + Compressed Impurity (CC) mechanisms, see
equations~(\ref{HR})--(\ref{CC}) in Sec.~\ref{sec_steps}.
 
For the $\Sigma 5$ grain boundary containing 0.5~ML of Bi, which
is not yet embrittled, we observe that removal of host
atoms from pure boundary sites that will become occupied by Bi (the
HR mechanism) strongly dominates and almost fully accounts for
the total decrease of $\Wsep$ in table~\ref{tbl_steps}.
The effect of mechanical distortion (the SS mechanism) is modest
and is similar in magnitude to the effect of insertion of Bi back
into a vacant site---the CC mechanism. This mechanism always acts
so as to strengthen the boundary.  This is exactly what one would
expect from Ag but perhaps not from Bi if one subscribed to the
``electronic'' point of view.  Since the Compressed
Impurity contribution can only decrease $\Wsep$, yet CC is
positive, it is clear that the Chemical mechanism is a large
effect tending to increase the work of separation. Even though
the grain boundary \textbf{C} is pre-streched, the fact that
replacing a vacancy even with Bi increases the intergranular
cohesion is yet another demonstration that the {\it electronic}
grain boundary weakening mechanism of embrittlement can be safely
ruled out. {\it The introduction of Bi divorced from any other
effect does not weaken the boundary---it strengthens it.}

Increasing Bi coverage to 1.0~ML results in a much larger
expansion of the grain boundary. The SS contribution increases
substantially giving almost half of the overall decrease of
$\Wsep$. The HR term still dominates, since twice as many Cu
atoms are removed from the grain boundary plane. Correspondingly,
the positive CC contribution also increases, but the sum of all
three is large enough to render the grain boundary brittle. We
therefore conclude that Bi embrittlement of the Cu grain boundary
may be attributed to both the Host Removal and Substitutional
Structure mechanisms with the relative weight of the latter
increasing as the coverage increases.
 
The effect of Na is strikingly similar to that of Bi
(table~\ref{tbl_steps}), especially for 1 ML coverage where not
only the total effect but even the contributions of each
mechanism are almost identical. Hence everything said about Bi
can be transferred to Na. At 0.5~ML, the fact that sodium atoms
are ``softer'' but ``larger'' (see table~\ref{tbl_fcc}) becomes
noticeable with some interesting consequences. The grain boundary
expansion and hence the SS contribution is smaller but the total
decrease of $\Wsep$ is larger for Na than for Bi. This is because
the CC contribution is almost zero. It would not be correct to
conclude that the bonding between Na and Cu is very weak for the
effect of CC at 1 ML is as large as for Bi. This is a perfect
demonstration of the effect of the Compressed Impurity mechanism
which combines with the Chemical contribution having opposite
sign and nearly cancels it out.

Silver does not embrittle Cu grain boundaries, and indeed we
observe that the HR and CC contributions strongly dominate but
almost cancel each other (but see the caption to
table~\ref{tbl_steps}). At 0.5~ML the ``loose'' site is so
attractive for a slightly larger Ag atom that it binds there more
strongly to other host atoms than even Cu itself does, with
negligible grain boundary expansion. At 1~ML silver has to
replace Cu at ``tight'' sites. As a result, HR~+~CC changes sign
and the grain boundary expands. These combine to decrease
$\Wsep$ by very small amount which is insufficient to lead to
embrittlement.

\section{Summary and conclusions}
\label{sec_concl}

We have investigated {\it atomic level\/} mechanisms in the
embrittlement of the $\Sigma 5$(310)[001] grain boundary in Cu by
Bi atoms. To this end, we did first principles calculations of
the grain boundary with up to 1~ML of impurity atoms: Bi, Na, and
Ag. We had two principal aims.
\begin{enumerate}
\setlength{\itemsep}{0in}
\setlength{\parsep}{0in} \setlength{\parskip}{0in} \setlength{\parindent}{0in}
\item Our first was to settle the controversy concerning whether
embrittlement is a {\it structural\/} or an {\it electronic}
effect. We can reconcile our earlier claim in
ref.~[\onlinecite{rainer04}] for a structural effect in the
$\Sigma 19$a grain boundary with apparently contradictory results
for the $\Sigma 5$ grain boundary in ref.~[\onlinecite{ruhle04}]
by new calculations of the latter which serve to reinforce the
conclusions reached in ref.~[\onlinecite{rainer04}].
We believe the grain boundary studied in
ref.~[\onlinecite{ruhle04}] was not in fact embrittled.

In our view the notion that metallic bonds may be either weakened
or strengthened by charge transfer is not tenable. Confusion has
arisen because calculations supporting these arguments were done
using {\it clusters} of atoms,\cite{MB,EV} the effects of
screening by the metallic electron gas were thereby
excluded;\cite{SP,Goodwin2} furthermore total energies were not
calculated.  We do not on the other hand reject chemical effects,
but as we have shown these {\it strengthen} not
weaken a grain boundary, at least in the case of substitutional
impurities and when defined as here with respect to insertion
into a vacant site as opposed to replacement by a host atom.

\item Our second aim was to unfold the complex character of
embrittlement by dividing the phenomenon into independent
constituents defined so that for any ``cause of embrittlement''
there exists a satisfactory thought experiment that isolates
it. We then proceed to assess in terms of our numerical data
which of these constituents plays the more prominent role.
\end{enumerate}

Our main conclusions can be summarised as follows.  
\begin{enumerate}
\setlength{\itemsep}{0in}
\setlength{\parsep}{0in} \setlength{\parskip}{0in}
\item 
Segregation of Bi to the $\Sigma 5$ Cu grain boundary leads to a
reduction in its work of separation. We predict that at least a
monolayer will segregate resulting in a ductile to brittle transition.
\item
We also predict Na to be nearly as good an embrittler as Bi. A
striking similarity between the embrittling propensity of free
electron metal Na and semi metallic Bi is strong evidence that
the embrittlement is primarily a structural effect.
\item
Ag is shown not to
embrittle the grain boundary, in agreement with experiment. In
fact at 0.5~ML it acts as a modest cohesion enhancer at the
$\Sigma 5$ grain boundary. This is really a size effect: the
larger Ag atom fits well into the ``loose site'' at the $\Sigma
5$ grain boundary. 
\item
We suggest an unambiguous, if not unique, thought experiment
applicable to both substitutional and interstitial impurities,
which allows one to quantify contributions to the changes in
$\Wsep$ coming from the following sources: ({\it i\/}) depletion
in the interface region (surface or grain boundary) of host atoms
(for substitutional impurities only); ({\it ii\/}) stretching of
the host--host bonds at the interface;
and ({\it iii\/}) insertion of the impurity into an empty site of 
pre-stretched interface. 

We find the first mechanism  
to be mostly responsible for the reduction of the work of separation of the
boundary with 0.5 ML of Bi or Na. At 1 ML the boundary becomes brittle     
due to the combined action of the first two mechanisms, whereas the third
mechanism acts in the opposite sense.

\end{enumerate}

\section*{Acknowledgement}

We would like to thank Rainer Schweinfest for discussions and
unpublished results. AYL has benefited from discussions with Ruth
Lynden-Bell, Pietro Ballone, and Ali Alavi. We acknowledge
Malachy Montgomery for computer system management. The work was
supported by EPSRC and Invest Northern Ireland.

\vskip -12pt


\end{document}